\documentclass[twocolumn,aps,prd,floatfix,nofootinbib]{revtex4-1}

\usepackage{graphicx,amsmath,amssymb,bm}
 \usepackage[colorlinks,pdfstartview=FitH]{hyperref}
 \usepackage{comment}
 \hypersetup{linkcolor=blue,citecolor=blue,filecolor=black,urlcolor=blue}

\newcommand\dy{\Delta y}
\newcommand\dyp{(\dy)}
\newcommand\dyc{\dy_{\rm corr}}
\newcommand\la{\langle}
\newcommand\ra{\rangle}

\begin{document}

\title {Acceptance dependence of fluctuation measures near the QCD critical point}

\author{Bo Ling}
\affiliation{Physics Department, University of Illinois at Chicago, Chicago, 
Illinois 60607, USA}
\author{Mikhail A.~Stephanov} 
\affiliation{Physics Department, University of Illinois at Chicago, Chicago, 
Illinois 60607, USA}



\date{December 2015}

\begin{abstract}
  We argue that a crucial determinant of the acceptance dependence of fluctuation measures in heavy-ion collisions is the range of correlations in the momentum space, e.g., in rapidity, $\Delta y_{\rm corr}$. The value of $\Delta y_{\rm corr}\sim1$ for critical thermal fluctuations is determined by the thermal rapidity spread of the particles at freezeout, and has little to do with position space correlations, even near the critical point where the spatial correlation length $\xi$ becomes as large as $2-3$ fm (this is in contrast to the magnitudes of the cumulants, which are sensitive to $\xi$).  When the acceptance window is large, $\Delta y\gg\Delta y_{\rm corr}$, the cumulants of a given particle multiplicity, $\kappa_k$, scale linearly with $\Delta y$, or mean multiplicity in acceptance, $\langle N\rangle$, and cumulant ratios are acceptance independent. While in the opposite regime, $\Delta y\ll\Delta y_{\rm corr}$, the factorial cumulants, $\hat\kappa_k$, scale as $(\Delta y)^k$, or $\langle N\rangle^k$. We demonstrate this general behavior quantitatively in a model for critical point fluctuations, which also shows that the dependence on transverse momentum acceptance is very significant. We conclude that extension of rapidity coverage proposed by STAR should significantly increase the magnitude of the critical point fluctuation signatures.
\end{abstract}


\maketitle

\section{Introduction}

Mapping the QCD phase diagram is one of the most important goals of
the heavy-ion collision experiments. A prominent feature on this map
of the thermodynamic states of QCD is the critical point punctuating the
first order phase transition between hadron matter and quark-gluon
plasma phase. Although this scenario is suggested by many models of
QCD thermodynamics as well as some lattice calculations, the precise
location (in temperature vs baryochemical potential, $T\mu_B$, plane) and even
the existence of this point is an open question, which so far has
eluded attempts to answer it using theoretical tools, such as
first-principle lattice simulations (for reviews, see, e.g.,
Refs.~\cite{Stephanov:2004wx,Schmidt:2006us,Stephanov:2007fk,Gupta:2009mu,Philipsen:2009dn,Schmidt:2009qq,Li:2010dya,Fukushima:2010bq,Aarts:2013lcm}).

The approach pursued by experiments to discover the critical point is
based on the analysis of the event-by-event
fluctuations \cite{Stephanov:1998dy,Stephanov:1999zu,Koch:2008ia}.  In 
the thermodynamic limit the
critical point is a thermodynamic singularity, where the intensive
measures of fluctuations violate the central limit theorem and
diverge. In a realistic heavy-ion collision this divergence is cut off
by the interplay of finite-time, non-equilibrium effects and the effect
of the critical slowing
down\cite{Stephanov:1999zu,Berdnikov:1999ph,Son:2004iv}. 
If, by varying the collision energy $\sqrt s$, one can create
fireballs with freezeout conditions close to the critical point, one
expects to observe non-monotonous dependence of fluctuation measures
on $\sqrt s$ as the critical point is approached and then passed.  The
search for the critical point using such beam energy scan strategy is
underway at the Relativistic Heavy Ion Collider (RHIC) at the
Brookhaven National Laboratory (BNL) and at the Super Proton
Synchrotron (SPS) at CERN in
Geneva~\cite{Stephans:2008zz,Mohanty:2009vb,Odyniec:2013aaa,Schuster:2009ak,Grebieszkow:2012vj}.

In order to compare experimental measurements with theoretical
predictions, as well as the results of different experiments to each
other, it is essential to understand the dependence of the fluctuation
measures on the size of the detector acceptance window, which varies
among experiments, or even different analyses of the same
experiment. The goal of the paper is to elucidate and quantify
this dependence.

The focus of this work is on critical point signatures. In particular,
on the higher-order cumulants sensitive to the thermodynamic
conditions at freezeout~\cite{Ejiri:2005wq,Karsch:2010ck} and
especially to the critical
fluctuations~\cite{Stephanov:2008qz}. However, we
begin with a more general analysis of acceptance dependence, which we
then illustrate using the critical point fluctuations. The purpose of
this paper is to examine the physics behind this dependence and
demonstrate it in a simple, analytic, but quantitatively realistic
model. Our qualitative arguments and quantitative results complement
and extend analyses of the acceptance dependence of the critical
fluctuations in Refs.\cite{Stephanov:2001zj,Stephanov:2004wx} to
higher-order cumulants, and also complement and contrast the analyses
of
Refs.\cite{Pruneau:2002yf,Bzdak:2012ab,Sakaida:2014pya,Kitazawa:2014nja,Kitazawa:2015ira,Asakawa:2015ybt,Garg:2013ata,Karsch:2015zna}
of non-critical correlations (see also
reviews~\cite{Jeon:2003gk,Koch:2008ia} for further references).

The type of questions we wish to answer are, for example,
what is the effect on critical point signatures of
increasing the transverse momentum range, say, 
from $p_T\in(0.4,0.8)$ GeV to $(0.4,1.2)$ GeV, as has been done in
the recent analysis by STAR \cite{Luo:2015ewa}? Or, 
what is the effect of extending the rapidity window from
$\dy=1$ to $\dy=1.5$ for protons, which will result from upgrading the inner
sectors of the Time Projection Chamber (iTPC) proposed~\cite{STAR-iTPC:2015} by the STAR experiment
at RHIC?


\section{Acceptance dependence of cumulants}
\label{sec:accept-depend-cumul}

Our main goal is to provide a transparent description
of the acceptance dependence of fluctuation measures, which can be
used to build quantitatively precise tools necessary to extract
physics from experimental data. To prevent the complexity of the
heavy-ion collision from obscuring the relevant features we wish to
highlight, we start with the
simplest idealized Bjorken model \cite{Bjorken:1982qr} of a boost invariant fireball 
and consider the dependence of the
fluctuation measures on the rapidity acceptance window $\Delta
y$. This will allow us to gain understanding of the main characteristics of the
acceptance dependence, which we can then carry to a more realistic
model with transverse expansion and $p_T$ acceptance dependence. 

Let us denote the number, or multiplicity, of accepted particles of a
given species (e.g., protons or pions) by $N$.\footnote{The acceptance
  dependence of the fluctuations of a particle number as opposed to, e.g.,
  net-proton or net-charge, is considerably more transparent and
  allows us to focus on the most important features.} The mean over
all events, $\la N\ra$, is then proportional to $\dy$ due to the boost
invariance:
\begin{equation}
  \label{eq:<N>}
  \la N \ra \sim \dy.
\end{equation}

How do the cumulants of order $k$, $\kappa_k[N]$, of $N$ vary with $\dy$? The
answer crucially depends on the range of the correlations in rapidity,
which we denote by $\dy_{\rm corr}$. 

Different contributions to the correlations (initial conditions, 
HBT, thermal/hydro fluctuations, critical fluctuations, etc.) are
characterized by different $\dy_{\rm corr}$. In this paper we shall
focus on critical point fluctuations, but we begin with a more general
discussion of correlations and their effect on acceptance dependence.

It is important to distinguish two qualitatively different regimes:
 $\dy \gg \dy_{\rm corr}$ and  $\dy \ll \dy_{\rm corr}$.

When $\dy \gg \dy_{\rm corr}$, all cumulants grow linearly
with $\dy$, because uncorrelated contributions are additive, by
construction, in a cumulant.
It is convenient and customary to remove this trivial volume dependence by
normalizing cumulants by their trivial, uncorrelated (Poisson) value ($\la N\ra$, for
cumulants of $N$), defining
\begin{equation}\label{eq:omega}
\omega_k\equiv\frac{\kappa_k}{\la N\ra}.
\end{equation}
The contribution of physical (e.g., critical)
correlations to this quantity,
$(\omega_k-1)$, 
saturates at a constant value for
$\dy\gg\dyc$.\footnote{Eq.~(\ref{eq:omega}) is not the only natural 
way to normalize the cumulant. Another widely used normalization is
$\kappa_k/\kappa_2$. Since, in practice, $\kappa_2-\protect\la N\protect\ra\ll \protect\la N\protect\ra$,
there is little difference between this normalization and Eq.~(\ref{eq:omega}).}

In the opposite regime, $\dy\ll\dy_{\rm
  corr}$, 
since the cumulants approach Poisson distribution values in the limit
$\Delta y\sim \la N\ra\to0$, we shall focus on the deviation of the
cumulants from their Poisson value, $\kappa_k-\la N\ra$. It is
convenient to express $\kappa_k-\la N\ra$ as a linear combination of
factorial cumulants, $\hat\kappa_l$, of equal or lower orders 
\begin{equation}
  \label{eq:kkhatS}
  \kappa_k-\la N\ra = \hat\kappa_k + \sum_{l=2}^{k-1} S(k,l)\hat\kappa_l
\end{equation}
where $S(n,m)$ are Stirling numbers of the second kind.  The most useful
property of the factorial cumulants is that each $\hat\kappa_k$
 measures the strength of the
(connected) $k$-particle correlation, and is therefore proportional to
the number of correlated $k$-plets, which scales roughly as $N^k$,
i.e., $\dyp^k$. This property is known (see, e.g., Ref.{\cite{Tannenbaum:1995dd}} 
and references therein), but for completeness and to provide a better intuitive understanding we derive it for
$k\le4$ in Appendix~\ref{sec:very-small-accept}.

Because of the simple asymptotic behavior of the factorial cumulants in
both regimes of $\dy$:
\begin{equation}
  \label{eq:tkgg-1}
  \hat \kappa_k \sim \dy \sim \la N\ra
\qquad(\dy\gg\dyc) 
\end{equation}
and
\begin{equation}
  \label{eq:tkll-1}
  \hat \kappa_k \sim \dyp^k \sim \la N\ra^k
\qquad(\dy\ll\dyc) .
\end{equation}
it is more convenient to describe acceptance
dependence in terms of the factorial cumulants $\hat\kappa_k$.

In contrast, the behavior of the normal cumulants, $\kappa_k$, in the
regime $\dy\ll\dyc$ is more complicated. According to
Eqs.~(\ref{eq:kkhatS}) and~(\ref{eq:tkll-1}), the limit $\dy\to0$ is controlled by the
lowest cumulant, $\hat\kappa_2$, i.e.,
\begin{equation}\label{eq:k-Ndy2}
\kappa_k-\la N\ra \sim\hat\kappa_2\sim\dyp^2 \qquad\mbox{when } \dy\to0 ,
\end{equation}
or $\omega_k-1\sim\dy$.\footnote{This helps explain the
  linear dependence of the normalized cumulants as the acceptance $\Delta\to0$ in
  Ref.\cite{Kitazawa:2015ira}. 
  We thank M.~Kitazawa for a discussion
  of this point.}
On the other hand, if or when the approximate hierarchy
$|\hat\kappa_k|\gg|\hat\kappa_l|$ for $k>l$ holds, as experimental
results \cite{Luo:2015ewa} indicate at some energies (e.g., at $\sqrt
s=7.7$ GeV
$|\hat\kappa_4|\gg|\hat\kappa_3|,|\hat\kappa_2|$ for $\dy\sim1$), the scaling in the regime $\dy\ll\dyc$, but not too small,
could be dominated by the highest cumulant in Eq.~(\ref{eq:kkhatS}),
and then
\begin{equation}\label{eq:k-Ndyk}
\kappa_k-\la N\ra \sim\hat\kappa_k\sim\dyp^k \qquad\mbox{($\dy$
  not too small)} ,
\end{equation}
or $\omega_k-1\sim \dyp^{k-1}$. The crossover between this behavior
and that in
Eq.~(\ref{eq:k-Ndy2}) could be a source of  non-monotonous
acceptance dependence of $\omega_k$ in some cases.

\section{Critical point correlations}
\label{sec:crit-point-corr}

In order to describe the acceptance
dependence of the fluctuations measures (the cumulants) 
more quantitatively we need to
input the physical information about the correlations. We shall focus
on {\em critical point} contribution to the fluctuations and use the model
described in Ref.~\cite{Stephanov:1999zu,Stephanov:2001zj} and, in application to higher-order
cumulants, in Refs.~\cite{Stephanov:2008qz,Stephanov:2011pb,Athanasiou:2010kw}. 
In this model the multiplicity
fluctuations at freezeout near the critical point receive a
contribution due to the coupling of the critical mode~$\sigma$ -- a collective
mode of fluctuations whose correlation length~$\xi$ becomes large (and
diverges at the critical point in the theoretical limit of infinitely
large system size and lifetime).

\subsection{The range of correlations}
\label{sec:range-correlations}

\begin{figure}[h]
  \centering
  \includegraphics[width=.32\textwidth]{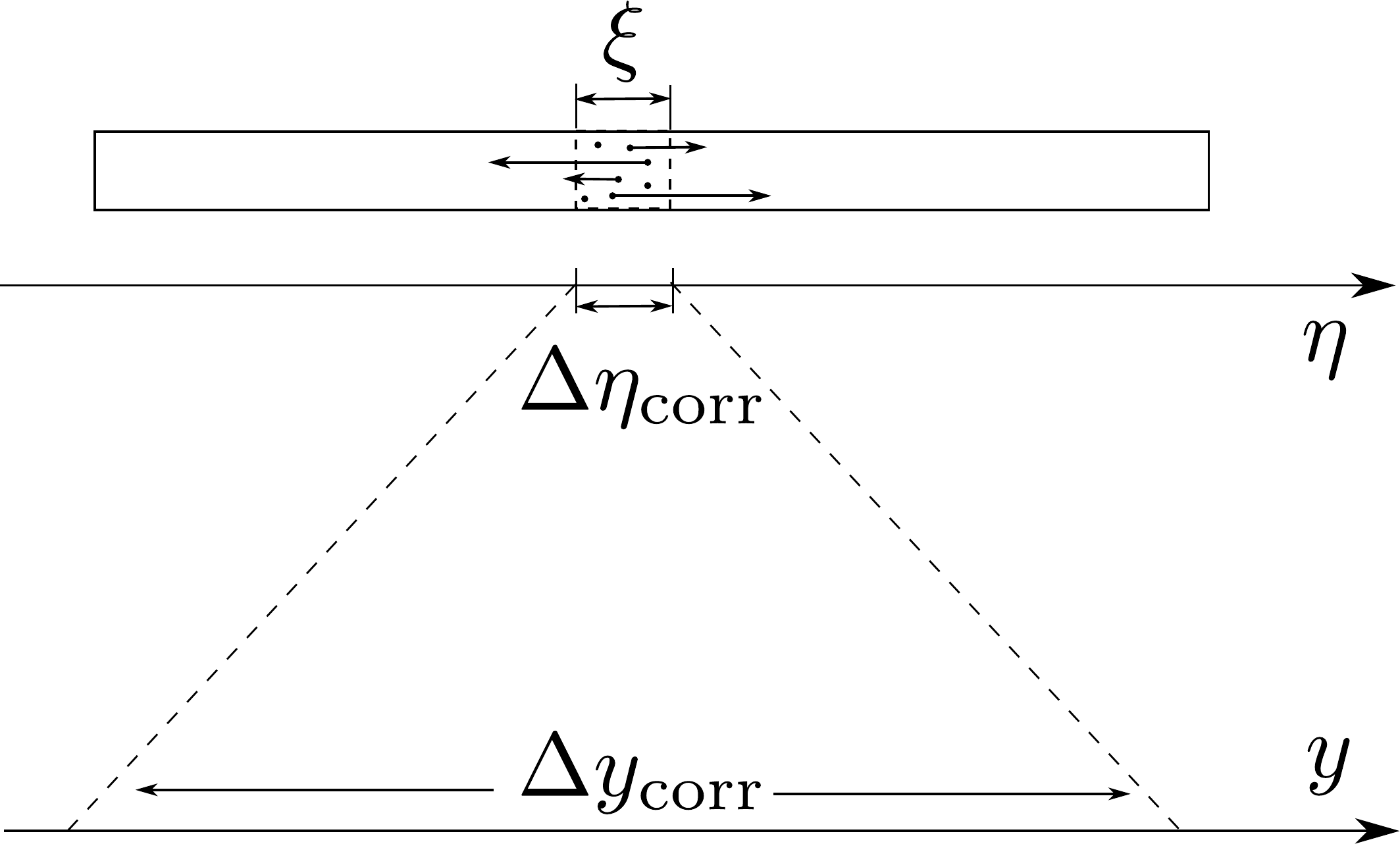}
\caption[]{\label{fig:deta-dy} Schematic illustration of the relation
  between the spatial (Bjorken) rapidity $\eta$ 
and kinematic rapidity $y$ via the effect of the thermal broadening (freezeout smearing).}
\end{figure}

What determines $\dy_{\rm corr}$? This depends on the physics behind
the correlations, and in the case we consider, the critical point, it
is the fluctuating collective mode.  Consider the boost-invariant
scenario with the correlation length in co-moving coordinates at
freezeout given by~$\xi$ (Fig.~\ref{fig:deta-dy}). This translates
into Bjorken rapidity correlation length $\Delta \eta_{\rm corr}
\approx\xi/\tau_{\rm f}$. With~$\xi$ ranging from 1~fm typically to
about $2-3$ fm near the critical point~\cite{Berdnikov:1999ph} and
with freezeout Bjorken time $\tau_{\rm f}\sim 10$ fm one estimates
$\Delta\eta_{\rm corr}\sim 0.1-0.3$.

\begin{figure}[h]
  \centering
\vskip 1em
  \includegraphics[height=9em]{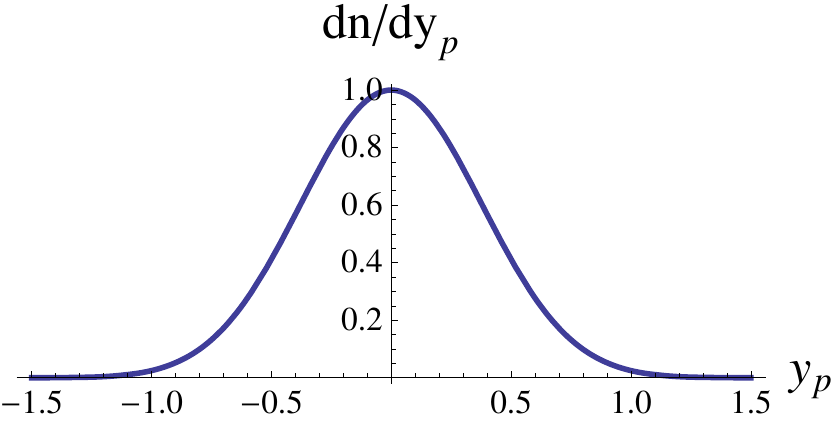}
  \caption[]{\label{fig:yp}Thermal proton rapidity distribution
    ($T=160$ MeV).}
\end{figure}

Detectors, however, do not measure the position space (Bjorken)
rapidity $\eta$, but the kinematic rapidity $y$ of the
particles. Within the spatial correlation volume $\Delta\eta_{\rm
  corr}$ at freezeout thermal distribution of particle rapidities
$y_p$ in the co-moving frame ranges roughly from $-1$ to $1$
(Fig.~\ref{fig:yp}). The observed rapidity $y=\eta+y_p$ of the
particles from each correlated volume is then spread over an interval
of order $\Delta y_{\rm corr}\sim 1$ (Fig.~\ref{fig:deta-dy}). Such
thermal broadening, or freezeout smearing, in the translation of
hydrodynamic spatial correlations into kinematic correlations has been
discussed recently in, e.g., Refs.\cite{Kapusta:2011gt,Ling:2013ksb}.
Because $\dy_{\rm corr}\gg\Delta\eta_{\rm corr}$, the value of
$\dy_{\rm corr}$ is not sensitive to~$\xi$. This is in contrast to the
magnitude of $\omega_k-1$~\cite{Stephanov:2008qz} --- larger~$\xi$
means more correlated particles in the same $\dy_{\rm corr}$ and
larger value of $\omega_k-1$.

It is essential for this argument that, within the correlated spatial
volume, particles of all momenta in the thermal distribution are
correlated with each other, as they are in the case of the critical
point fluctuations we consider --- see Eq.~(\ref{eq:dfdf}) below.

\subsection{The model of critical correlations}
\label{sec:model-crit-corr}

To make calculations simpler and the results more transparent, we
shall use the observation that even the maximal correlation length
$\xi\lesssim2-3$ fm (limited in a heavy-ion collision by finite-time
and critical slowing down effects~\cite{Berdnikov:1999ph}) is still
considerably smaller than the typical size of the system. A measure of
that size is either the transverse radius, or the Bjorken proper time
at freezeout, $R\sim7-10$ fm. In the idealized limit $R\gg\xi$ we can
consider the spatial correlation as almost local and approximate it by
a delta function in the integrals involving slowly (in position space)
varying distribution functions $f(\bm x,\bm p)$. The normalization of
the delta function is fixed by matching the space integral of the
correlator:
\begin{equation}
  \label{eq:ss}
\int d^3\bm x \la\sigma(\bm x)\sigma(\bm y)\ra = T\xi^2
~\Rightarrow~    \la\sigma(\bm x)\sigma(\bm y)\ra \to
T\xi^2\delta^3(\bm x-\bm y)
\end{equation}
Similar approach can be applied to the 3-point and 4-point (connected)
functions,
which can be also approximated by delta-functions, normalized using the
results of Ref.\cite{Stephanov:2008qz}:
\begin{equation}
  \label{eq:sss}
  \la\sigma(\bm x)\sigma(\bm y)\sigma(\bm z)\ra \to
-2\tilde\lambda_3T^{3/2}\xi^{9/2}\delta^6(\bm x,\bm y,\bm z)
\end{equation}
where $\delta^6(\bm x,\bm y,\bm z)\equiv\delta^3(\bm x-\bm
y)\delta^3(\bm x-\bm z)$ and
\begin{equation}
  \label{eq:ssss}
  \la\sigma(\bm x)\sigma(\bm y)\sigma(\bm z)\sigma(\bm w)\ra_c
\to
 6(2\tilde\lambda_3^2-\tilde\lambda_4)
 T^2 \xi^7\delta^9(\bm x,\bm y,\bm z,\bm w)
\end{equation}
where $\delta^9(\bm x,\bm y,\bm z,\bm w)\equiv\delta^3(\bm x-\bm
y)\delta^3(\bm x-\bm z)\delta^3(\bm x-\bm w)$.
The parameters $\tilde\lambda_3$ and $\tilde\lambda_4$ are dimensionless
functions of $T$ and $\mu_B$ characterizing the non-gaussianity of the
fluctuations of $\sigma$ and described in more detail in
Ref.\cite{Stephanov:2008qz}.  We shall not be concerned with the {\em
  absolute} magnitude of (the critical contribution to) the cumulants
in this paper, and therefore the normalization factors, such as $\tilde\lambda_3$
and $\tilde\lambda_4$ will not be essential in the following. 

The contribution of the critical mode to the fluctuation of the
particle distribution function, $f_A$, is given by~\cite{Stephanov:2011pb}
\begin{equation}
  \label{eq:df}
  (\delta f_A)_\sigma = -\frac{\chi_A}{\gamma_A}g\sigma(\bm x_A), 
\end{equation}
where we introduced a shorthand for the set of phase-space coordinates,
e.g., $A=(\bm x_A,\bm p_A,\ldots)$ with ''$\ldots$'' standing for any additional
particle quantum numbers (spin, charge, etc.). We denoted the
derivative of the equilibrium distribution 
function,
 $f^{\rm eq}_A=(\exp(\epsilon_A-\mu)/T)\pm1)^{-1}$, by
\begin{equation}
  \label{eq:chip}
  \chi_A \equiv \frac{\partial f^{\rm eq}_A}{\partial\mu}
= \frac1T f^{\rm eq}_A(1\pm f^{\rm eq}_A),
\end{equation}
(plus/minus for fermions/bosons), where the coupling $g$ is defined as derivative of the effective mass of the
particle in a given background of the critical mode: 
$g\equiv dm(\sigma)/d\sigma$. For protons this corresponds to
effective sigma-model coupling $g\sigma\bar\Psi\Psi$. We also denoted
by $\gamma_A
= d\epsilon_A/dm=m/\epsilon_A$ the relativistic gamma-factor of the particle
with momentum $\bm p_A$.

Using Eq.~(\ref{eq:df})
 we can calculate correlators of these critical fluctuations.
For example, for the critical contribution, denoted by
$\la\ldots\ra_\sigma$, to the two-point correlator we find 
\begin{equation}
  \label{eq:dfdf}
  \la \delta f_A \delta f_B \ra_\sigma = 
T\xi^2g^2
\frac{\chi_A}{\gamma_A}\frac{\chi_B}{\gamma_B}
\delta^3(\bm x_A-\bm x_B)
\end{equation}
Note that the correlation in Eq.~(\ref{eq:dfdf}) appears local in the
position space (on scales $R\gg\xi$), but is non-local in the
momentum space. This momentum space non-locality is essential for the
argument in Section~\ref{sec:range-correlations}.

Introducing the shorthand notation, $\int_A$, for the integration over
the phase space, i.e., over $\bm x_A$, $\bm p_A$ (as well as the
summation over spin) we can express the fluctuation of the particle
number $N$ in terms of $\delta f$:  $\delta N = \int_A \delta f_A$.
The critical contribution to the 
quadratic cumulant of the fluctuations is then given by
\begin{equation}
  \label{eq:k2}
  \kappa_2[N]_\sigma\equiv\la (\delta N)^2\ra_\sigma
=\int_{A}\int_{B}{T}\xi^2{g^2}
\frac{\chi_A}{\gamma_A}\frac{\chi_B}{\gamma_B}
\delta^3(\bm x_A-\bm x_B)
\end{equation}
One of the spatial integrals (say, over $\bm x_B$) can be performed
using the delta-function. The result can be written as
\begin{equation}
  \label{eq:k2-x}
  \kappa_2[N]_\sigma\equiv\la (\delta N)^2\ra_\sigma
=\int_{\bm x} \ {T}\xi^2{g^2}
 \left(\int_{\hat A} \frac{\chi_A}{\gamma_A}\right)^2
\end{equation}
where we introduced the shorthand, $\int_{\hat A}$, for the integration 
over the momentum (and summation over spin) subspace of the phase
space $A=(\bm x,\bm p_A, \ldots) = (\bm x, \hat A)$. The dependence on
the acceptance enters via the range of the momentum integration in
$\int_{\hat A}$. This range is fixed in the frame of the detector
(lab frame), but translates (by boost) 
into different ranges in the co-moving frame for different points
on the freezeout surface.

Treating higher-order cumulants similarly we find
\begin{equation}
  \label{eq:k3-x}
  \kappa_3[N]_\sigma\equiv\la (\delta N)^3\ra_\sigma
=\int_{\bm x} \ 2\tilde\lambda_3 T^{3/2} \xi^{9/2} g^3
 \left(\int_{\hat A} \frac{\chi_A}{\gamma_A}\right)^3
\end{equation}
and 
\begin{multline}
  \label{eq:k4-x}
  \kappa_4[N]_\sigma
\equiv\la (\delta N)^4\ra_\sigma - 3\la (\delta N)^2\ra_\sigma^2
\\=\int_{\bm x} \ 6\left(2\tilde\lambda_3^2-\tilde\lambda_3\right)
 T^2 \xi^7 g^4
 \left(\int_{\hat A} \frac{\chi_A}{\gamma_A}\right)^4
\end{multline}

Our model of the critical fluctuations is not precise or detailed
enough to distinguish between factorial, $\hat\kappa_k$, and normal
cumulants $\kappa_k-\la N\ra$. However, it is clear from the physical
origin ($k$-particle correlation) and from the small $\dy$ behavior of
$\kappa_k[N]_\sigma$, that they describe contributions to the
factorial cumulants $\hat\kappa_k$.  However, the model describes the
regime of sufficiently large correlation length $\xi$, when critical
contributions are proportional to higher powers of $\xi$ for higher
cumulants, $\hat\kappa_k\sim \xi^{5k/2-3}$ \cite{Stephanov:2008qz},
thus leading to the approximate hierarchy
$|\hat\kappa_4|\gg|\hat\kappa_3|\gg|\hat\kappa_2|$. 

If such hierarchy does
not hold in the data (in particular, when $\hat\kappa_4$ is
close to zero because it changes sign~\cite{Stephanov:2011pb}, or when
$\dy$ is very small), one
should then directly compare $\kappa_k[N]_\sigma$ to experimentally measured
factorial cumulants, $\hat\kappa_k$, instead of $\kappa_k-\la N\ra$.

\subsection{Transverse expansion}
\label{sec:transverse-expansion}

We shall use the blast wave
model (see, e.g., Refs.~\cite{Schnedermann:1993ws,Retiere:2003kf} and, in
application to transverse momentum correlations, Refs.~\cite{Voloshin:2003ud,Gavin:2008ev,Moschelli:2009tg,Ling:2013ksb})
 of the freezeout surface to perform
the integrals in Eqs.~(\ref{eq:k2-x})--(\ref{eq:k4-x}). In this
model the freezeout surface is
isochronous, at a given Bjorken time $\tau=\tau_{\rm f}$, and the 4-velocity field
is given, in longitudinal Bjorken coordinates $(\tau,\eta)$ and
transverse polar coordinates $(r,\phi)$ as
\begin{equation}
  \label{eq:u-}
  (u^\tau,u^\eta,u^r,u^\phi) = (\cosh \eta_\perp,0,\sinh \eta_\perp,0)
\end{equation}
where the transverse velocity is parameterized as
\begin{equation}
  \label{eq:kappa}
 \beta_\perp= \tanh \eta_\perp \equiv \frac{u^r}{u^\tau} = \beta_s 
\frac{r}{R}
\end{equation}
where the surface velocity
 $\beta_s\approx 0.6$ is a parameter we determine by fitting the inclusive
single-particle distribution.

The space integral in Eqs.~(\ref{eq:k2-x})--(\ref{eq:k4-x}) becomes the integral over
the freezeout hypersurface:
\begin{equation}
  \label{eq:x-eta}
  \int_{\bm x} \to \tau \int d\eta \int r dr \int d\phi
\end{equation}
The momentum integration in Eqs.~(\ref{eq:k2-x})--(\ref{eq:k4-x}) can be expressed as an
integral over kinematic rapidity, transverse momentum $p_\perp$ and
azimuthal angle $\psi$:
\begin{multline}
  \label{eq:Ahat}
  \int_{\hat A}\frac{1}{\gamma_A}\to
d_{\rm s} 
\int d^3\!\left(\frac{\bm p}{2\pi}\right)\frac{1}{\gamma_{\bm p}}
\\
\to   \frac{d_{\rm s}m}{(2\pi)^3}\int_{y_1}^{y_2} dy \int_{p_{\rm min}}^{p_{\rm max}} p_\perp dp_\perp \int_0^{2\pi} d\psi
\end{multline}
where $d_s$ is the species degeneracy (e.g., $d_{\rm s}=2$ for proton
spin). The acceptance cuts are represented by the limits of the
integrations, with $y_2-y_1\equiv \Delta y$.
The particle distribution function is given in terms of the energy
$\epsilon_{\bm p}=u\cdot p$ in the co-moving frame and expressed in
terms of the lab frame $y$, $p_\perp$ and $\psi$, as well as $\eta$,
$r$ and $\phi$, via
\begin{equation}
  \label{eq:udotp}
  u\cdot p = 
m_\perp\cosh(y-\eta)\cosh \eta_\perp
- p_\perp\cos(\psi-\phi)\sinh \eta_\perp,
\end{equation}
where $m_\perp\equiv\sqrt{m^2+p_\perp^2}$ and $\eta_\perp$ is given by 
Eq.~(\ref{eq:kappa}). Boltzmann approximation for the equilibrium
distribution function
\begin{equation}
  \label{eq:f-Bol}
  f^{\rm eq} \approx \exp\left(\frac{\mu - u\cdot p}{T}\right),
\qquad
\chi \approx \frac {f^{\rm eq}}{T}.
\end{equation}
is sufficient for our purposes (in particular, for protons, assuming
$m-\mu_B\gg T$), and it allows performing part of the multiple
integration in Eqs.~(\ref{eq:k2-x})--(\ref{eq:k4-x}) analytically.


\subsection{Beyond the model}
\label{sec:beyond-model}

Before proceeding to discuss the results let us emphasize again that
one of our simplifying approximations is that of Bjorken
(boost) invariance of the freezeout hypersurface. This assumption does
not affect the validity of Eqs.~(\ref{eq:k2-x})--(\ref{eq:k4-x}). The formulas are more general, and can be applied also for a realistic
freezeout surface obtained in a hydrodynamic simulation, as it is done
in Ref.\cite{Jiang:2015hri}. The purpose of our paper is to address the issue of the
acceptance dependence in a transparent fashion. The use of the Bjorken
scenario (with a blast-model transverse flow) allows us to separate
the acceptance window size dependence from, e.g., the dependence on
the location of the acceptance window (central vs forward rapidity),
which has a completely different physical origin.

The approximation $R\gg\xi$, as in, e.g.,
Eq.~(\ref{eq:ss}), is also very helpful to simplify our treatment, but
can be relaxed, if necessary, along the lines of
Refs.\cite{Stephanov:2009ra,Jiang:2015hri}. To be consistent, however,
this should be accompanied by the inclusion of finite-time,
non-equilibrium effects.

We need to keep in mind that the model we use to
demonstrate the acceptance dependence is the most
basic model of critical fluctuations, which neglects 
non-equilibrium effects except for one -- the critical slowing down effect
limiting the magnitude of $\xi$. Nevertheless, the 
model should be sufficient to describe qualitatively and semiquantitatively
the acceptance dependence of the cumulants, largely because this dependence is
constrained by generic considerations described in
Section~\ref{sec:accept-depend-cumul}.

The most important feature of this dependence is~$\dyc$, which is
determined by the thermal momentum distribution of the particles at
freezeout, and is not much sensitive to the dynamics of the spatial
correlations.  The feature of the critical fluctuations which is very
important for both $\dy$ and $p_T$ window dependence of the
fluctuations is the non-locality of the correlations in the {\em
  momentum} space, e.g., in Eq.~(\ref{eq:dfdf}). We do not expect that
taking into account non-equilibrium effects more thoroughly (along the
lines of, e.g., Ref.\cite{Stephanov:2009ra} or
\cite{Mukherjee:2015swa}) will affect this property significantly.

We do wish, however, to underscore the importance of developing a more
comprehensive non-equilibrium approach to fluctuations 
to enable more quantitative
comparison with experiment~\cite{Nahrgang:2015tva}. 
Such an approach is especially crucial for
predicting the absolute magnitudes of the cumulants (which sensitively
depend on $\xi$ \cite{Stephanov:2008qz}) as well as their sign \cite{Asakawa:2009aj,Stephanov:2011pb,Mukherjee:2015swa}.

\section{Results}
\label{sec:results}

We can now use the formulas we derived for critical point contributions
to (factorial) cumulants to predict the acceptance dependence of these
contributions. We choose $\sqrt s=19.6$ GeV as a representative 
collision energy. The results are very similar at other energies we
considered (e.g., 7.7 and 11.5 GeV) and in agreement with the general
arguments described in Section~\ref{sec:accept-depend-cumul}.

We determine the temperature and chemical potential at freezeout using the
fit from Ref. \cite{Cleymans:2005xv}:
$T\approx 160$ MeV, $\mu_B \approx 200$ MeV.

\begin{figure}\centering
  
  \includegraphics[width=0.4\textwidth]{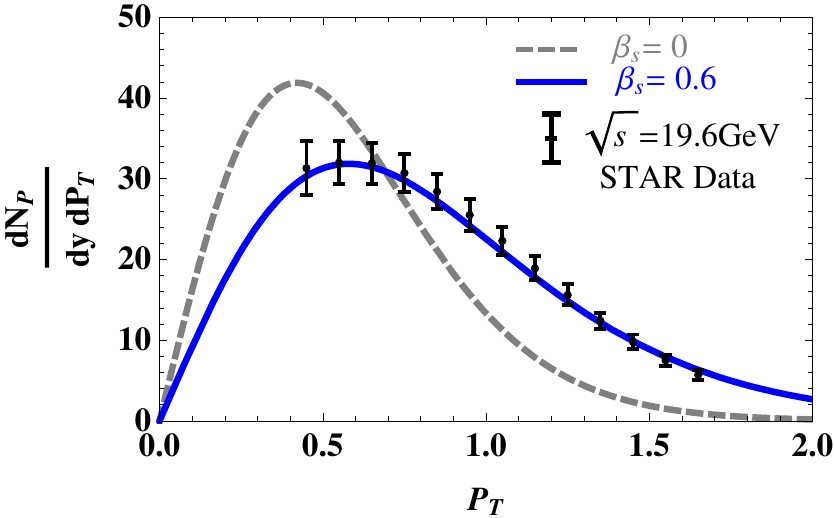}
  
  \caption{Spectrum of proton transverse momenta from experiment (points)~\cite{Kumar:2014tca},
    static thermal distribution (dashed line) and blast-wave model
    (solid line). $p_T$ is measured in GeV.}
\label{fig:spectrum}
\end{figure}

We use the value of $\beta_s$ (the radial surface velocity) optimizing the
agreement with the proton $p_T$ spectrum, as shown in Fig.~\ref{fig:spectrum}.

We normalize the proton cumulants $\kappa_k[N]$ by their
Poisson value, $\la N\ra$, as in Eq.~(\ref{eq:omega}) and consider
 the contribution of critical
fluctuations:
\begin{equation}\label{eq:oks}
\omega_{k,\sigma} = \frac{\kappa_k[N]_\sigma}{\la N\ra}.
\end{equation}
to  $\omega_k-1$ or, more precisely, to $\hat\kappa_k/\la N\ra$ (see
discussion at the end of Section~\ref{sec:model-crit-corr}).

\begin{figure}\centering
  \includegraphics[width=0.4\textwidth]{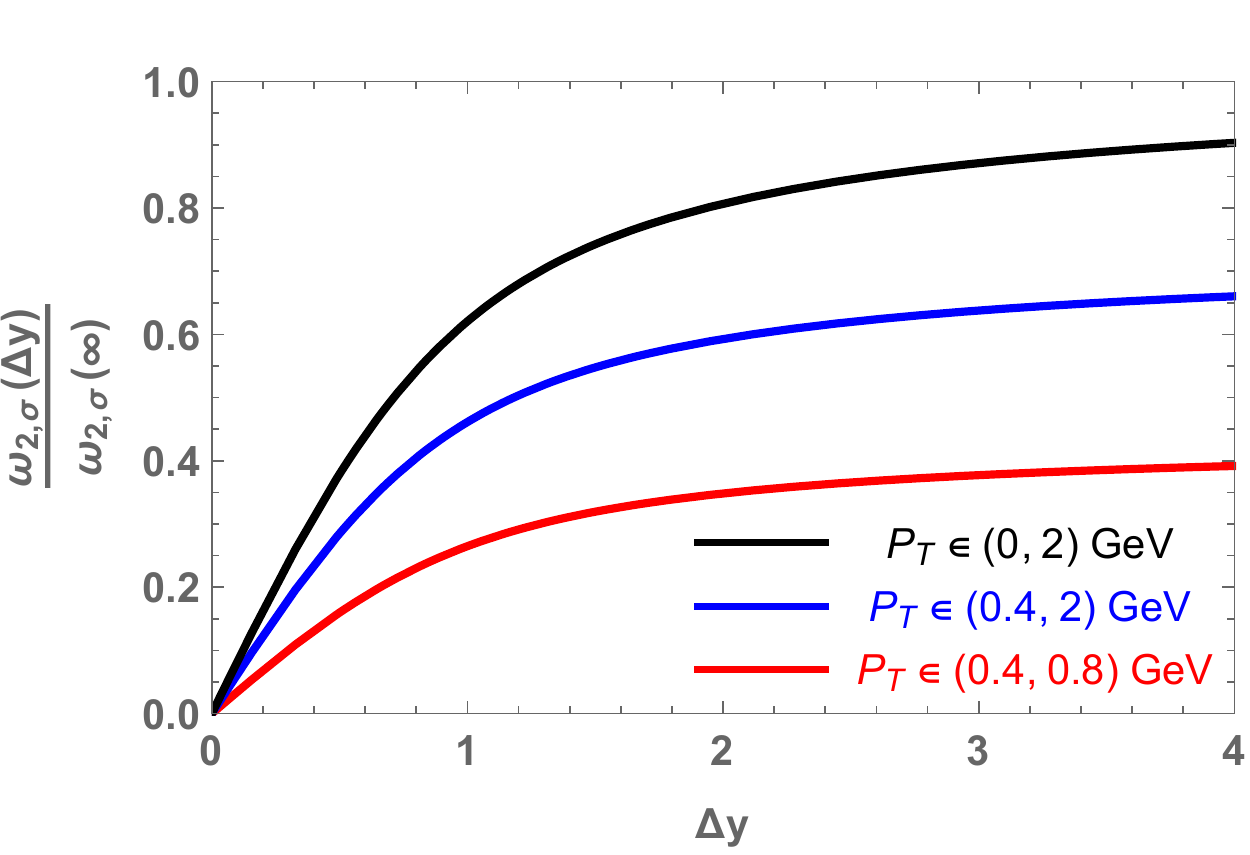}

  \includegraphics[width=0.4\textwidth]{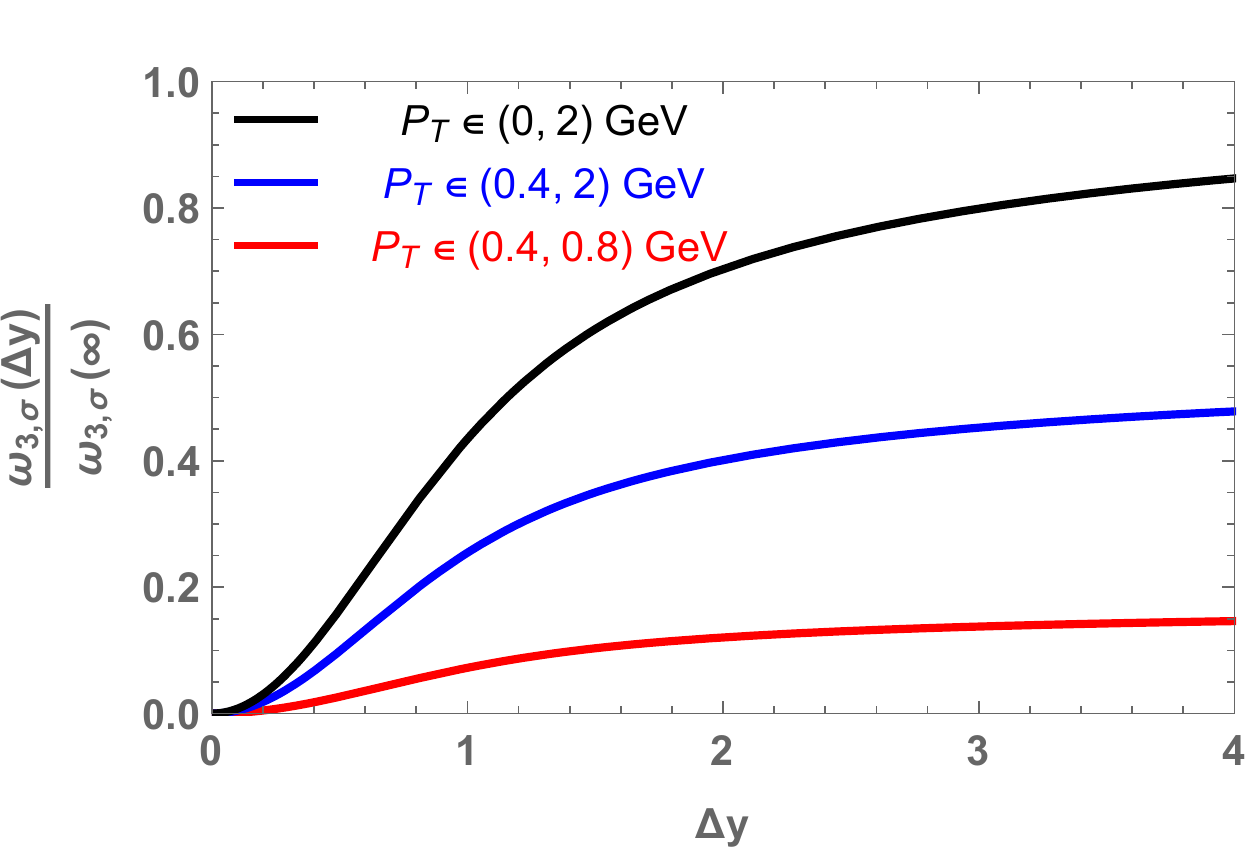}

  \includegraphics[width=0.4\textwidth]{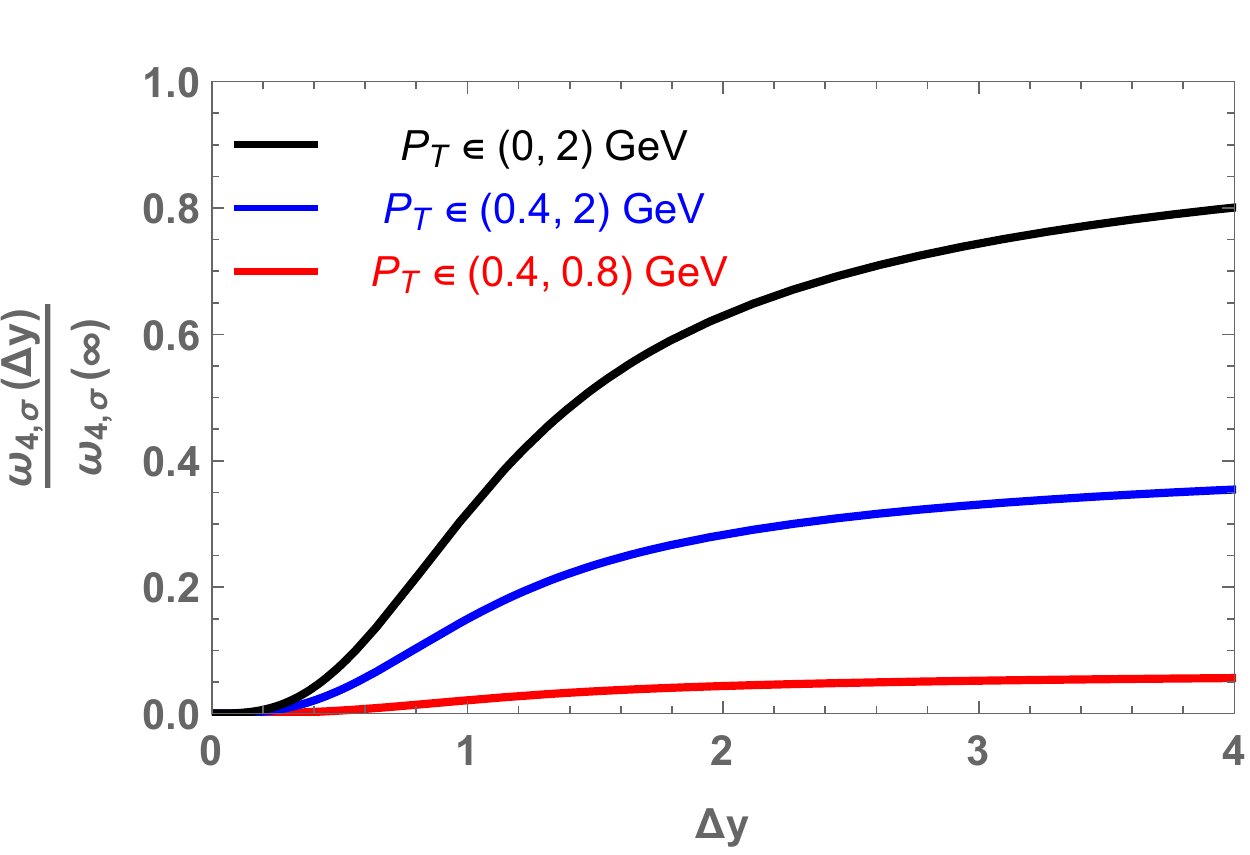}

  \caption[]{Acceptance dependence of the critical contribution to the
    normalized cumulants of proton number. See Section~\ref{sec:results}.}
\label{fig:o234}
\end{figure}

This quantity depends on acceptance window and, as expected from the
arguments in Section~\ref{sec:accept-depend-cumul}, 
saturates in the limit of full (infinite in
rapidity $y$ and transverse momentum $p_T$) acceptance at a value we
denote $\omega_{k,\sigma}(\infty)$. To show the acceptance dependence
we plot the ratio of $\omega_{k,\sigma}$ in the given acceptance
window ($\Delta y$ in rapidity for three representative sets of $p_T$
cuts) to the full acceptance value $\omega_{k,\sigma}(\Delta
y)/\omega_{k,\sigma}(\infty)$ in Fig.~\ref{fig:o234}.
In this ratio the prefactors such as $g$, $\xi$ and $\tilde\lambda_i$
in Eqs.~(\ref{eq:k2-x})--(\ref{eq:k4-x}) cancel.

To understand the origin of the $\Delta y$ scale at which the
dependence saturates, it is helpful to look at the thermal rapidity
distribution of protons shown in Fig.~\ref{fig:yp}.
The width of this distribution essentially sets the scale of the rapidity
smearing of correlations (schematically pictured in Fig.~\ref{fig:deta-dy}),
and the corresponding $\Delta y_{\rm corr}\sim 1$. 

For small $\Delta y \ll 1$, the expected behavior
$\omega_{k,\sigma}(\dy)\sim(\Delta y)^{k-1}$ comes, in the model, from
the factor $(\int_{\hat A}\chi_A/\gamma_A)^k\sim\dyp^k$ in
Eqs.~(\ref{eq:k2-x})--(\ref{eq:k4-x}) where the
volume of the integration domain  scales as $\Delta y$ (see
Eq.~(\ref{eq:Ahat})), and from division by $\la N\ra\sim\dy$ in
Eq.~(\ref{eq:oks}).

Fig.~\ref{fig:o234} also demonstrates that the $p_T$ window dependence
is significant, especially, for higher-order cumulants. This is a
simple consequence of the non-locality of the particle correlations in
{\em momentum} space as seen in, e.g., Eq.~(\ref{eq:dfdf}). In other
words, particles of all momenta in the thermal distribution 
are correlated with each other by critical fluctuations.

\section{Summary and conclusions}
\label{sec:conclusions-outlook}

We have described the basic features of the acceptance dependence of
the fluctuation measures, in particular, of the (factorial) cumulants of the
proton number fluctuations. The main lesson from our analysis is 
that dependence on rapidity acceptance window $\Delta y$ is determined
by the correlation range $\Delta y_{\rm corr}$ in momentum space which
has very little to do with the correlation length in Bjorken
coordinate rapidity $\eta$. The latter is related to the spatial
correlation length $\xi$: $\Delta \eta_{\rm corr}\sim \xi/\tau_{\rm f}$ and is
typically negligible compared to the former, $\Delta y_{\rm corr}\sim 1$, which is
due to the thermal distribution of the particles in the kinematic
rapidity.
The value $\Delta y_{\rm corr}$ separates two regimes of rapidity
window dependence. 

For $\dy\gg \dy_{\rm corr}$ the cumulants grow linearly with 
$\dy$ and their ratios, such as, e.g, $\kappa_k[N]/\la N\ra=\omega_k$,
approach constant values. 

The opposite, small acceptance regime,  $\dy\ll \dy_{\rm corr}$,
is easier to describe using the {\em factorial} cumulants,   $\hat\kappa_k$,
because they scale as $\dyp^k$ in this regime.
The normal cumulants, on the other hand, given by linear
combinations of factorial cumulants in Eq.~(\ref{eq:kkhatS}), 
 have a more complicated,
polynomial dependence on $\dy$ in this regime.

Because the factorial cumulants have much simpler scaling in both
large and small acceptance regimes
(Eqs.~(\ref{eq:tkgg-1}),~(\ref{eq:tkll-1})), we conclude that using these
cumulants to analyze the acceptance dependence is advantageous.

For the typical experimental acceptance, $\Delta
 y \lesssim 1$, we find that 
larger acceptance leads to significantly larger
critical point signal, especially for higher-order cumulants of
fluctuations (see Fig.~\ref{fig:o234}). Larger $p_T$ acceptance has a
 similar effect.\footnote{  
 It is important to also note that the statistical error of the
 estimator of the cumulants grows {\em slower} with acceptance in this
 regime: as
 $\langle N\rangle^{k/2}$, or $\dyp^{k/2}$. This result is implicit in
 the analysis of Ref.\cite{Luo:2014rea}, where the error is estimated by $(\kappa_2/\epsilon)^{k/2}\mathcal N_{\rm ev}^{-1/2}$, with $\mathcal N_{\rm ev}$ the number of events and $\epsilon$ the
 detector efficiency, since,  in practice, $\kappa_2\approx\la N\ra$.}
These results underscore the importance of the
 planned STAR detector iTPC upgrade~\cite{STAR-iTPC:2015} to extend
 the rapidity coverage for the critical point search in the
 Beam-Energy Scan experiment at RHIC.

\acknowledgments

One of the authors (M.S.) would like to thank Berndt Muller, Nu Xu,
and ZhangBu Xu for initiating the discussion which lead to
this paper, as well as the participants and the
organizers of the EMMI Workshop ``Fluctuations in Strongly Interacting
Hot and Dense Matter: Theory and Experiment'' for illuminating
discussions. 
This material is based upon work supported by the U.S. Department of
Energy, Office of Science, Office of Nuclear Physics under Award
Number DE-FG0201ER41195.

\newpage

\bibliography{Cumulants}


\appendix

\section{Small acceptance and factorial cumulants}
\label{sec:very-small-accept}

For completeness, we provide a derivation of the claim,
used in Section~\ref{sec:accept-depend-cumul},
that the $k$-th factorial cumulant, $\hat\kappa_k$,
measures the strength of the (connected) $k$-particle correlation and,
therefore, in the regime $\dy\ll\dyc$, roughly counts the number of
correlated $k$-plets, leading to $\hat\kappa_k\sim \dyp^k$.  We shall 
start with a 2-point correlator and build up to
derive Eq.~(\ref{eq:kkhatS}) up to $k=4$
in order to elucidate the relationship between normal and
factorial cumulants.

Let us choose an infinitesimally small parameter
$\varepsilon$ and divide a given kinematic region into $\mathcal O(1/\varepsilon)$
infinitesimally small cells, or bins, labeled by index $a$. 
We shall denote by $n_a$ the random, fluctuating event-by-event,
occupation number of bin $a$, and by $\la n_{a} \ra$ 
its event average. When $\varepsilon\to0$ the value of $\la n_{a}\ra={\cal O}(\varepsilon)\ll 1$ and thus $n_{a}$
obeys Poisson statistics with infinitesimally small mean.
In other words, the probabilities are given by
\begin{multline}
  \label{eq:P01}
  P_{n_a=0} = 1 - \la n_a\ra + \mathcal O(\varepsilon^2),
~\\
  P_{n_a=1} = \la n_a\ra + \mathcal O(\varepsilon^2),
~
P_{n_a\ge2} = \mathcal O(\varepsilon^2).
\end{multline}
i.e., most of the time $n_a=0$, very seldom $=1$, and almost never
$\ge2$.  This means, in particular, that in expectation values we can
replace $n_a^2$ with $n_a$: $\la n_a^2 \ldots \ra = \la n_a \ldots \ra
\times \left(1 + \mathcal O(\varepsilon^2)\right)$. Similarly, we can
also derive the following equation for the fluctuation $\delta
n_a=n_a-\la n_a\ra$ which we shall find useful:
 \begin{equation}
   \label{eq:dnkn}
   \la (\delta n_{a})^k\ldots\ra=\la n_{a}\ldots\ra 
\times \left(1 + \mathcal O(\varepsilon^2)\right),
\qquad (k\ge2)
 \end{equation}

Using this equation we can obtain the following expression for the
2-point
correlator:
\begin{equation}
  \label{eq:nn}
  \la \delta n_{a}  \delta n_{b}\ra 
= \la n_a\ra\delta_{ab} + C_{ab}
\end{equation}
The first term on  the r.h.s., nonzero only when $a=b$, is
simply the contribution of the fluctuation of the number of particles
in a given bin~$a$. It does not represent correlations. All
correlations are in the second term, $C_{ab}$, which is
nonzero when $a\neq b$. The feature of this term important
for us is
that in the limit $\varepsilon\to 0$, $C_{ab}$ varies very little
within the acceptance window if $\dy\ll\dyc$, by definition of $\dyc$.

Let us sum in Eq.~(\ref{eq:nn}) over all the bins
within the acceptance window. By definition,
$\sum_a n_a = N$, and thus
\begin{equation}
  \label{eq:NN}
\kappa_2\equiv
  \la (\delta N)^2\ra = \sum_a \la n_a\ra + \sum_{ab} C_{ab} = \la N\ra + \hat\kappa_2
\end{equation}
The first term on the r.h.s.\  is the well-known result of
Poisson statistics. The last term is the contribution of
correlations. When the acceptance window is much smaller than the
range of the correlations, $\dy\ll\dyc$, this term 
is proportional to the volume of the
acceptance window squared, $\dyp^2$ or $\la N\ra^2$, because in this case
we can approximate  $C_{ab}$ by a constant within
the acceptance window.\footnote{
Note that $C_{ab}={\cal O}(\varepsilon^2)$, but it cannot be
neglected compared to the first, $\mathcal O(\varepsilon)$, term in Eq.~(\ref{eq:nn}) because in
Eq.~(\ref{eq:NN}) the
number of elements in the sum $\sum_a$ is ${\cal
  O}(1/\varepsilon)$, while in $\sum_{ab}$ it is ${\cal
  O}(1/\varepsilon^2)$, so the two terms in Eq.~(\ref{eq:NN}) are of
the same order, $\mathcal O(\varepsilon^0)$, finite in the limit
$\varepsilon\to 0$. On the other hand, the diagonal 
terms, $C_{aa}$, in Eq.~(\ref{eq:nn}) are negligible,
since their contribution to Eq.~(\ref{eq:NN}) is ${\cal
  O}(\varepsilon)\to 0$.}

Let us generalize this argument to the 3-particle correlator:
\begin{multline}
  \label{eq:nnn}
  \la \delta n_a  \delta n_b \delta n_c\ra 
= \la n_a\ra\delta_{ab}\delta_{ac} 
\\+ (\delta_{ab} C_{ac} + \delta_{ac} C_{ab} + \delta_{ab} C_{ac}) 
+ C_{abc}
\end{multline}

\begin{figure}\centering
  \includegraphics[scale=.18]{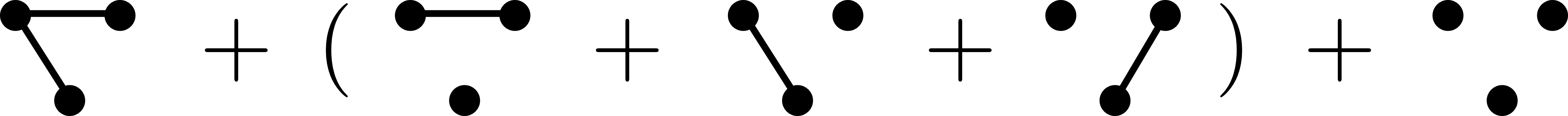}
  \caption{Diagrammatic representation of the r.h.s. of Eq.~(\ref{eq:nnn})}
\label{fig:3dots}
\end{figure}

The 5 terms on the r.h.s. can be represented by diagrams
as shown in Fig.~\ref{fig:3dots}, where a dot represents a bin and a line connecting
two dots represents two bins which coincide, due to, e.g.,
$\delta_{ab}$. The first term is nonzero only when $a=b=c$ and then it
equals $\la (\delta n_a)^3 \ra = \la n_a\ra$ due to Eq.~(\ref{eq:dnkn}).
The next 3 terms (in the parentheses) are nonzero only when 2 of the 3 bins
coincide. For example, when $a=b\neq c$, they give $\la (\delta n_a)^2 \delta n_c\ra
=\la \delta n_a\delta n_c\ra=C_{ac}$. 

Summing over the bins within the acceptance we find
\begin{multline}
  \label{eq:NNN}
\kappa_3\equiv
  \la(\delta N)^3\ra = \sum_a \la n_a\ra + 3\sum_{ab}C_{ab}
  +\sum_{abc}C_{abc} \\
= \la N\ra + 3\hat\kappa_2 + \hat\kappa_3
\end{multline}
It is easy to see that in the regime when the size of the acceptance
is much larger than the range of correlations, $\dy\gg\dy_{\rm corr}$
 each term in the
r.h.s. scales with the volume of the acceptance window, $\Delta y$, or
$\la N\ra$. While in the opposite regime, small acceptance window
$\dy\ll\dy_{\rm corr}$, due to the smoothness of $C_{abc}$, the term
$\hat\kappa_3$ scales as volume to the power 3, or $\dyp^3$.

\begin{figure}\centering
  \includegraphics[scale=.18]{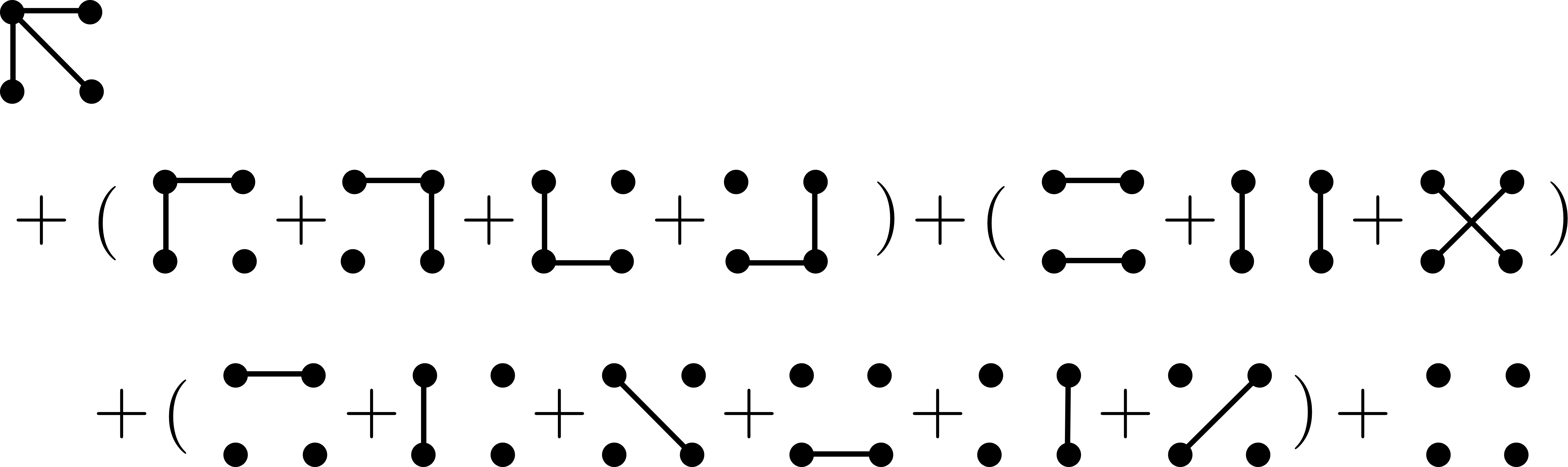}
  \caption{Diagrammatic representation of the r.h.s. of Eq.~(\ref{eq:nnnn_c})}
\label{fig:4dots}
\end{figure}

Defining connected 4-point correlator as usual:
\begin{multline}
  \label{eq:nnnn_c-def}
    \la \delta n_a  \delta n_b \delta n_c \delta n_d\ra_c
\equiv \la \delta n_a  \delta n_b \delta n_c \delta n_d\ra
\\- (\la \delta n_a  \delta n_b\ra\la \delta n_c \delta n_d\ra
+ b\leftrightarrow c +  b\leftrightarrow d)
\end{multline}
we can express it as
\begin{multline}
  \label{eq:nnnn_c}
  \la \delta n_a  \delta n_b \delta n_c \delta n_d\ra_c
=
 \la n_a\ra\delta_{ab}\delta_{ac}\delta_{ad} 
\\+ (\delta_{ab} \delta_{ac} C_{ad} + \mbox{ 3 more}) 
+ (\delta_{ab} \delta_{cd} C_{ac} + \mbox{ 2 more}) 
\\ + (\delta_{ab}  C_{acd} + \mbox{5 more terms}) 
+ C_{abcd}
\end{multline}
We have written only one of the similar terms 
in each set of the parentheses. The additional
terms are easier to represent diagrammatically, as shown in Fig.~\ref{fig:4dots}.

Summing over all bins we obtain
\begin{multline}
  \label{eq:NNNN}
\kappa_4\equiv  
\la(\delta N)^4\ra_c = \la(\delta N)^4\ra -(\la(\delta N)^2\ra)^2
\\=
 \sum_a\la n_a\ra + 7\sum_{ab}C_{ab}
  +6\sum_{abc}C_{abc} +\sum_{abcd}C_{abcd}
\\= 
\la N\ra + 7\hat\kappa_2 + 6\hat\kappa_3 + \hat\kappa_4
\end{multline}

The quantities we denoted by $\hat\kappa_k$, and defined so far as $k$-fold
sums of corresponding $C$'s, can be recognized as
factorial cumulants of the random variable $N$ (with
$\hat\kappa_1=\kappa_1=\la N\ra$). In the same way that normal
cumulants (for $k>2$) measure the deviations from the normal
distribution, the factorial cumulants (for $k>1$) 
measure deviations from the Poisson distribution.

 We find that the
scaling of the factorial cumulants
with the acceptance window volume, or with $\dy$, when
this window is very small, $\dy\ll \dyc$, is given by Eq.~(\ref{eq:tkll-1}), while 
for $\dy\gg\dyc$ the scaling is the same as for the normal cumulants,
linear in $\dy$, Eq.~(\ref{eq:tkgg-1}).

In experiment, the factorial cumulants can be easily calculated from the
measured cumulants by solving Eqs.~(\ref{eq:NN}),~(\ref{eq:NNN})
and~(\ref{eq:NNNN}) for $\hat\kappa_k$:
\begin{multline}
  \label{eq:tk}
\hat\kappa_1=\kappa_1=\la N\ra,
\quad
\hat \kappa_2 = \kappa_2 -\kappa_1,
\quad\\
\hat \kappa_3 = \kappa_3 - 3\kappa_2 + 2\kappa_1,
\quad\\
\hat \kappa_4 = \kappa_4 - 6\kappa_3 + 11\kappa_2 - 6\kappa_1
\end{multline}
(the coefficients here are Stirling numbers of the first kind).

 Alternatively,
one can use
the expansion of the generating function:
\begin{equation}
  \label{eq:gen}
  g(x) \equiv \sum_{k=1}^\infty \hat\kappa_k \frac{x^k}{k!} = \ln \left\la (1+x)^{N}\right\ra.
\end{equation}
to express the factorial cumulants directly in terms of the plain
moments $\la N^k\ra$, or in terms of the factorial moments
$\hat\mu_k=\left\la N(N-1)\ldots(N-k+1)\right\ra$ using
\begin{equation}
  \label{eq:g2}
  g(x) = \ln\left(1+\sum_{k=1}^\infty\hat\mu_k \frac{x^k}{k!}\right).
\end{equation}

\end{document}